\documentclass[prd,eqsecnum,twocolumn,amsfonts,amssymb]{revtex4}

\usepackage{graphicx}

\usepackage{bm}
\usepackage{amsmath}

\newcommand{\beq}{\begin{equation}}
\newcommand{\eeq}{\end{equation}}
\newcommand{\beqs}{\begin{eqnarray}}
\newcommand{\eeqs}{\end{eqnarray}}
\newcommand{\lsim}{\mathrel{\raisebox{-
.6ex}{$\stackrel{\textstyle<}{\sim}$}}}
\newcommand{\gsim}{\mathrel{\raisebox{-
.6ex}{$\stackrel{\textstyle>}{\sim}$}}}

\begin{document}

\title{An Extra-Dimensional Model of Dark Matter} 

\author{Sudhakantha Girmohanta and Robert Shrock}

\affiliation{\ C. N. Yang Institute for Theoretical Physics and
Department of Physics and Astronomy, \\
Stony Brook University, Stony Brook, New York 11794, USA }

\begin{abstract}

We present a model for dark matter with extra spatial dimensions in
which Standard-Model (SM) fermions have localized wave functions. The
underlying gauge group is $G_{\rm SM} \otimes {\rm U}(1)_z$, and the
dark matter particle is a SM-singlet Dirac fermion, $\chi$, which is
charged under the ${\rm U}(1)_z$ gauge symmetry.  We show that the
conventional wisdom that the mass of a Dirac fermion is naturally at
the ultraviolet cutoff scale does not hold in this model.  We further
demonstrate that this model yields a dark matter relic abundance in
agreement with observation and discuss constraints from direct and
indirect searches for dark matter.  The dark matter particle interacts
weakly with matter and has negligibly small self-interactions. Very
good fits to data from cosmological observations and experimental dark
matter searches are obtained with $m_\chi$ in the multi-TeV range. A
discussion is given of observational signatures and experimental tests
of the model.

\end{abstract}

\maketitle


\section{Introduction}
\label{intro_section}

There is strong evidence for dark matter (DM) comprising approximately
85 \% of the matter in the universe \cite{pdg_dm,dmv}.
One possibility is that the dark matter is a weakly interacting
massive particle (WIMP) \cite{kolb_turner}-\cite{blanco}, entailing
associated new physics beyond the Standard
Model (BSM) \cite{pbh}.  Here we discuss a model of this type involving extra
spatial dimensions with fermion wave functions that are localized in
the extra dimensions. The underlying gauge group is
\beq
G = G_{\rm SM} \otimes {\rm U}(1)_z \ ,
\label{g}
\eeq 
where $G_{\rm SM} = {\rm SU}(3)_c \otimes {\rm SU}(2)_L \otimes {\rm
  U}(1)_Y$ is the Standard Model gauge group. We denote the gauge
fields for the hypercharge U(1)$_Y$ and the new U(1)$_z$ symmetries as
$B_\mu$ and $C_\mu$, respectively, and the corresponding field
strength tensors as $B_{\mu\nu} = \partial_\mu B_\nu - \partial_\nu
B_\mu$ and $C_{\mu\nu} = \partial_\mu C_\nu - \partial_\nu C_\mu$. The
gauge couplings of the SU(3)$_c$, SU(2)$_L$, U(1)$_Y$, and U(1)$_z$
gauge interactions are denoted $g_s$, $g$, $g'$, and $g_z$.  We show
that this model can produce the relic density of dark matter and also
satisfy other constraints from particle physics and cosmology.


\section{Basic Framework of the Model}
\label{model_section}

In this section we describe the model.  We make use of a theoretical
framework in which spacetime is taken to have $d=4+n$ dimensions, with
$n$ extra spatial dimensions. The usual spacetime coordinates are
denoted $x_\nu$, with $\nu=0,1,2,3$, and the $n$ extra spatial
coordinates are denoted $y_\lambda$, with $0 \le y_\lambda \le L$,
where $L$ is the compactification scale. Periodic boundary conditions
are assumed.  The Standard-Model fermion content is extended to
include three generations of electroweak-singlet neutrinos,
$\nu_{a,R}$, where $a=1,2,3$ is the generational index; we will refer
to this theory as the SM, with this extension of the fermion sector
being implicitly understood.  The fermions in the model are
(suppressing color and generational indices \cite{gennote})
\begin{widetext}
\beqs
&& Q_L = {u \choose d}_L: \ (3,2)_{1/3,z_Q}, \quad\quad 
u_R: \ (3,1)_{ 4/3,z_u}, \quad\quad
d_R: \ (3,1)_{-2/3,z_d} \cr\cr
&& L_L = {\nu_\ell \choose \ell}_L: \ (1,2)_{-1,z_L}, \quad\quad
\nu_{\ell,R}: \ (1,1)_{0,z_\nu}, \quad\quad
      \ell_R: \ (1,1)_{-2,z_\ell} \ , 
\label{fermions}
\eeqs
\end{widetext}
together with a Dirac dark matter fermion, $\chi$:
\beq
\chi_{L,R}: \ (1,1)_{0,z_\chi} \ ,
\label{chi}
\eeq
where the numbers in parentheses are the dimensionalities of the
fermion representations of the SU(3)$_c$ and SU(2)$_L$ factor groups
in $G$, and the subscripts are the weak hypercharge $Y_f$ and the
U(1)$_z$ charge, $z_f$ of the given fermion, $f$.  The U(1)$_z$ charge
of each type of SM fermion $f_{a,L}$ and $f_{a,R}$ is taken to
independent of the generation index $a$. The SM Higgs field is the
complex doublet $H={h^+ \choose h^0}$ transforming as $(1,2)_{1,z_H}$,
with vacuum expectation value (VEV)
$\langle H \rangle_0 = {0 \choose v/\sqrt{2} }$.  We recall that the
assignments of weak hypercharges to the SM fermions and Higgs fields
presume a normalization of the U(1)$_Y$ gauge interaction, since only
the products of hypercharges multiplied by the U(1)$_Y$ gauge coupling
$g'$ appear in covariant derivatives. Our normalization is indicated
by the usual relation for the electric charge operator, $Q_{em} =
T_{3L}+(Y/2)$ with the above-listed assignments of weak hypercharges
to SM fields.

Our model also includes a second Higgs field $\omega$ transforming as
$(1,1)_{0,z_{\omega}}$.  The U(1)$_z$ charges of $\omega$ and the
fermion fields are only defined up to an overall rescaling of the
U(1)$_z$ gauge coupling, $g_z$, since only the products of the $z$
charges multiplied by $g_z$ occur in covariant derivatives. We fix
this scale by setting $z_\omega=1$.  The potential terms for the
$\omega$ Higgs field are chosen such that the minimum occurs at
$\langle \omega \rangle_0 = v_\omega/\sqrt{2}$, spontaneously breaking
the U(1)$_z$ gauge symmetry.  We shall take $v_\omega \gg v$, i.e.,
the spontaneous breaking of the U(1)$_z$ symmetry occurs at a mass
scale that is much higher than the electroweak symmetry breaking scale
of $v \simeq 246$ GeV.  This is necessary in order for the model to be
consistent with precision electroweak data and with bounds from
collider searches for additional (neutral) vector bosons. The mass
generation and mixings for the neutral gauge bosons are discussed in
the appendix.  The coefficient of the cross term
$(H^\dagger H)(\omega^\dagger \omega)$ is assumed to be small enough so
that it does not significantly modify this pattern of electroweak and
U(1)$_z$ gauge symmetry breaking.

The wave function of each fermion $f$ has the form \cite{as,ms} form
\beq
\Psi_f(x,y)= \psi_f(x)\chi_f(y) \ .
\eeq
(Here, we follow our previous notation in \cite{nnb02}-\cite{nuled},
denoting the $y$-dependent part of $\Psi_f(x,y)$ as $\chi_f$; the distinction
with the dark matter fermion field $\chi$ is clear, since the latter has no
subscript $f$.)  The function $\chi_f(y)$ is localized at a point $y_f$ in the
extra dimensions, with a Gaussian profile
\beq
\chi_f(y) = A_f e^{-\mu^2 \|y-y_f\|^2} = Ae^{-\|\eta-\eta_f\|^2} \ , 
\label{chif}
\eeq
where $\|y_f\| = (\sum_{\lambda=1}^n y_{f,\lambda}^2)^{1/2}$ is the usual
Euclidean norm (defined with respect to periodic boundary conditions in the $n$
compact dimensions); $A_f$ is a normalization constant; and we define the
dimensionless variable
\beq
\eta_f \equiv \mu y_f \ .
\label{eta}
\eeq
The fermion localization length $\sigma \equiv 1/\mu$ satisfies
$\sigma \ll L$, indicating the strong localization of the fermion
wave functions in the extra dimensions. In terms of the dimensionless
quantity
\beq
\xi \equiv \mu L = \frac{L}{\sigma} \ ,
\label{xi}
\eeq
this localization means $\xi \gg 1$. 

We use a low-energy effective field theory (EFT) approach in which the
properties in the low-energy, long-distance 4D theory are calculated
by integrating over the short-distance compactified degrees of
freedom. The effective Lagrangians in $d=4+n$ and in $d=4$ dimensions
are denoted ${\cal L}_{{\rm eff},4+n}$ and ${\cal L}_{\rm eff}$,
respectively. The energy scale associated with the compactification is
defined as $\Lambda_L \equiv 1/L$. The model has an ultraviolet (UV)
cutoff, denoted $M_*$, with $M_* \gsim \mu \gg \Lambda_L$.  For
canonical normalization of fermion fields, one has
\beq
A_f = \Big ( \frac{2}{\pi} \Big )^{n/4}\, \mu^{n/2} \ .
\label{af}
\eeq
It will suffice here to discuss the lowest Kaluza-Klein (KK) modes;
effects of higher KK modes in this type of model are are discussed in
\cite{nuled,kk}.  The gauge and Higgs fields are taken to have flat
profiles in the extra dimensions.  An appeal of this type of
extra-dimensional model is that it can explain the hierarchy of
Standard Model (SM) fermion masses by appropriate placement of left-
and right-handed chiral components of SM fermions in the extra
dimensions \cite{as,ms}.  Owing to the different locations of these chiral
components of SM fermions, the model is often called a
``split-fermion'' theory.  The choices $\Lambda=10^2$ TeV, i.e., $L =
2 \times 10^{-19}$ cm, and $\xi = 30$, i.e., $\mu = 3 \times 10^3$
TeV, yields adequate fermion localization and enables the model to
account for quark and charged lepton masses and quark mixing, while
satisfying phenomenological constraints
\cite{as,ms,kk,bvd,nnb02,nuled,dif,ged}.  Bounds on proton decay are
satisfied by sufficiently large separation of quark and lepton wave
function centers in the extra dimensions \cite{as}. Interestingly,
$n-\bar n$ oscillations and associated dinucleon decays are not
suppressed and can occur at observable levels \cite{nnb02}.

For the case of $n=2$ extra dimensions, Ref. \cite{nuled} derived a
solution for charged lepton and neutrino wave function centers (with
three electroweak-singlet $\nu_{a,R}$ fields, $a=1,2,3$) that fits
data on neutrino masses and lepton mixing, as well as constraints from
flavor-changing neutral-current (FCNC) processes. We shall assume
$n=2$ here and again adopt the solution for fermion wave function
centers from \cite{nuled}.  Ref. \cite{nuled} considered two gauge
groups for SM fermions, namely $G_{\rm SM}$ and the left-right
symmetric group, $G_{\rm LRS} = {\rm SU}(3)_c \otimes {\rm SU}(2)_L
\otimes {\rm SU}(2)_R \otimes {\rm U}(1)_{B-L}$ \cite{lrs}, where $B$
and $L$ denote baryon and (total) lepton numbers.  Viable dark matter
candidates were presented for both of these gauge groups, each of
which involved a chiral dark matter fermion $\chi$ that is a nonsinglet under
the electroweak gauge symmetry.

Here we will present a different approach to dark matter in this
extra-dimensional framework, in which the dark matter particle is a
Dirac fermion transforming as a singlet under $G_{\rm SM}$ and vectorially
under the U(1)$_z$ gauge interaction, as indicated in (\ref{chi}). 

First, we show an important result, namely that this
model is a counterexample to conventional EFT model-building rules on Dirac
fermion masses.  This point is quite general and is independent of our
specific application to dark matter.  According to conventional EFT
lore, if a fermion can have a gauge-invariant Dirac mass term
$m_\chi \bar\chi\chi$, then the mass
$m_\chi$ is  generically be of order the UV cutoff and hence is 
integrated out of the low-energy EFT applicable below this cutoff.
However, our theory provides an example of how this conventional lore
can be misleading.  We show this for general $n$.  The key point is
that the mass term,
\beqs
&& {\cal L}_{{\rm eff},4+n,\chi} = m_{\chi,4+n}\bar\chi\chi \cr\cr
&=& m_{\chi,4+n} ( \bar\chi_L \chi_R + \bar\chi_R \chi_L) \ , 
\label{chibarchi}
\eeqs
involves the product
\beq
A_f^2e^{-(\|\eta-\eta_{\chi_L}\|^2+\|\eta-\eta_{\chi_R}\|^2)} \ . 
\label{chibarchiproduct}
\eeq
Integrating this over the extra coordinates, we get the 4D mass term
\beq
{\cal L}_{{\rm eff},m_\chi} = m_\chi \bar\psi_\chi(x)\psi_\chi(x) \ , 
\label{chibarchi4d}
\eeq
where
\beq
m_{\chi} = m_{\chi,4+n} \, e^{-(1/2)\|\eta_{\chi_L} - \eta_{\chi_R}\|^2} \ ,
\label{mchi4d}
\eeq
or equivalently 
\beq
\|\eta_{\chi_L} - \eta_{\chi_R} \| 
   = \Big [ 2\ln \Big ( \frac{m_{\chi,4+n}}{m_\chi} \Big ) \Big ]^{1/2} \ .
\label{mchrelgen}
\eeq
Even if one takes $m_{\chi,4+n}$ to be the largest mass scale in the theory, 
namely the UV cutoff, $M_*$, it is easy to separate wave function centers of
$\chi_L$ and $\chi_R$ in the higher dimensions sufficiently to get
$m_{\chi} \ll M_*$, owing to the Gaussian suppression factor in Eq.
(\ref{mchi4d}).  Explicitly, taking $m_{\chi,4+n}=M_*$, this distance is
\beq
\|\eta_{\chi_L} - \eta_{\chi_R} \| 
   = \Big [ 2\ln \Big ( \frac{M_*}{m_\chi} \Big ) \Big ]^{1/2} \ .
\label{mchrelmuv}
\eeq
As we will discuss below, a typical value of $m_\chi$ that produces a
dark matter relic density matching the observed value is $m_\chi = 10$
TeV.  Substituting this in Eq. (\ref{mchi4d})
together with the value that we take for the UV cutoff,
$M_* = 10\mu = 3 \times 10^4$ TeV, yields the corresponding modest 
separation distance $\|\eta_{\chi_L} - \eta_{\chi_R}
\| = [2\ln(3 \times 10^3)]^{1/2} = 4.0$.

This is an example of how naive dimensional analysis of operators in a
(4D) low-energy effective field theory may not capture all of the relevant
physics.  A well-known previous example of this is the
natural suppression of flavor-changing neutral current processes by the
Glashow-Iliopoulos-Maiani (GIM) mechanism \cite{gim}.  For example, an
operator contributing to $K^0 - \bar K^0$ mixing is the four-fermion operator
in the (4D) effective Lagrangian
\beq
    {\cal L}_{K-\bar K,{\rm eff}} = \frac{c^{(K \bar K)}}{\Lambda_{K \bar K}^2}
    [\bar s_L \gamma_\lambda d_L] [\bar s_L \gamma^\lambda d_L] + h.c.
\label{leff_kkbar}
\eeq
If one were to take $\Lambda_{K \bar K}$ to be a typical electroweak
symmetry-breaking scale, $\Lambda_{K \bar K} \simeq 250$ GeV, this
would lead to much too large a $K^0-\bar K^0$ mixing and hence much
too large a $K_L-K_S$ mass difference.  The solution to this problem
required the input of additional information about the theory at a
higher energy scale not included in the low-energy EFT, namely the
presence of the charm quark, filling out an SU(2)$_L$ doublet,
rendering the neutral weak current diagonal in mass eigenstates at
tree level \cite{gim} and also leading to the severe suppression of
FCNC processes such as $K^0 - \bar K^0$ mixing at the one-loop level
\cite{gl}. Similarly, an operator contributing to one-loop radiative
charged lepton flavor-violating (CLFV) decays of the form $\ell \to
\ell' \gamma$, such as $\mu \to e \gamma$, is
\begin{widetext}
\beq
    {\cal L}_{\ell \to \ell'\gamma,{\rm eff}} =
    \frac{1}{\Lambda_{CLFV}}
 \Big ( c^{(\ell'_L \ell_R)} [\bar \ell'_L \sigma_{\lambda \rho} \ell_R]
 + c^{(\ell'_R \ell_L)} [\bar \ell'_R \sigma_{\lambda \rho} \ell_L] \Big )
 F_{em}^{\lambda\rho} + h.c.
\label{leff_meg}
\eeq
\end{widetext}
Again, if one were to substitute a value of $\Lambda_{CLFV}$ of order
the electroweak symmetry breaking scale, $\Lambda_{CLFV} \simeq 250$
GeV, this would lead to an excessively large branching ratio $BR(\mu
\to e \gamma)$, in disagreement with experimental upper limits.  The
actual size of $\Lambda_{CLFV}$ depends on ultraviolet physics not
specified in the low-energy EFT, in particular on whether a theory
satisfies the conditions for natural suppression of lepton flavor
violation derived in \cite{leeshrock77}.  The present theory provides
a different, but analogous, example of how information about
ultraviolet physics must be added to the basic operator analysis in
the low-energy EFT.  In our case, this information on the UV physics
is given by the distances between wave function centers of the
relevant fermion fields in the extra dimensions.  The example here
involves a bilinear operator product that determines the mass of the
$\chi$ fermion, and in the discussion below we will see a similar
application to four-fermion operators that determine properties of the
dark matter in this model.


\section{Anomaly Constraints}
\label{anomaly_section}

We shall require that the 4D low-energy EFT is free of gauge
anomalies. Because $\chi$ is a SM-singlet, the anomaly cancellation
conditions (ACCs) involving just the factor groups of $G_{\rm SM}$ are
the same as in the Standard Model itself and hence are satisfied (independently
for each SM fermion generation).  These are the conditions that the
$[{\rm SU}(3)_c]^3$, $[{\rm SU}(3)_c]^2 \, {\rm U}(1)_Y$, $[{\rm
    SU}(2)_L]^2 \, {\rm U}(1)_Y$, and $[{\rm U}(1)_Y]^3$ gauge anomalies,
and the mixed gauge-gravitational anomaly, $(gr)^2 {\rm U}(1)_Y$
(where $gr=$ graviton), all vanish.  Since there are an even number of 
SU(2)$_L$ fermion doublets, namely $N_c+1=4$, there is also no global
SU(2)$_L$ anomaly. There are six new anomaly cancellation conditions.
Since $\chi$ is a Dirac fermion, it does not contribute to any gauge
anomalies. For simplicity, we shall assume that the ${\rm U}(1)_z$
charges of SM fermions are independent of the generational index. Then
the six new anomaly cancellation conditions are the following:
%
\beq
    [{\rm SU}(3)_c]^2 \, {\rm U}(1)_z: \  2z_Q - z_u - z_d = 0
\label{anom_su3su3u1z}
\eeq
%
\beq
    [{\rm SU}(2)_L]^2 \, {\rm U}(1)_z: \ 3z_Q +z_L=0
\label{anom_su2su2u1z}
\eeq
%
\beqs
[{\rm U}(1)_Y]^2 \, {\rm U}(1)_z: \ && z_Q + 3z_L - 8z_u - 2z_d - 6z_\ell  = 0
\cr\cr
&&
\label{anom_u1yu1yu1z}
\eeqs
%
\beqs
    {\rm U}(1)_Y \, [{\rm U}(1)_z]^2: \  && z_Q^2 - z_L^2 -2z_u^2 + z_d^2 +
    z_\ell^2  = 0 \cr\cr
&& 
\label{anom_u1yu1zu1z}
\eeqs
%
\beqs 
[{\rm U}(1)_z]^3: && \ 6z_Q^3-3z_u^3-3z_d^3+2z_L^3-z_\ell^3-z_\nu^3 = 0 \cr\cr
&&
\label{anom_u1zu1zu1z}
\eeqs
and
\beqs
(gr)^2 \, {\rm U}(1)_z: \ && 6z_Q - 3 z_u - 3z_d + 2z_L - z_\ell - z_\nu= 0 \ .
\cr\cr
&&
\label{anom_grgru1z}
\eeqs
A general solution of Eqs. (\ref{anom_su3su3u1z})-(\ref{anom_grgru1z}) was
presented in \cite{adh} (for a more general case in which there are an
arbitrary number of right-handed neutrino fields with different U(1)$_z$
charges). We briefly review this here, giving our own solution. To begin, one
observes that four of the six ACCs, namely Eqs. (\ref{anom_su3su3u1z}),
({\ref{anom_su2su2u1z}), (\ref{anom_u1yu1yu1z}), and (\ref{anom_grgru1z}), are
  linear in the six variables $z_Q$, $z_u$, $z_d$, $z_L$, $z_\nu$, and
  $z_\ell$. Thus, these four equations constitute a linear transformation $A:
  {\mathbb R}^6 \to {\mathbb R}^4$. Considering the above six $z$ charges as a
  vector $v = (z_Q,z_u,z_d,z_L,z_\nu,z_\ell)^T \in {\mathbb R}^6$, the four
  linear anomaly cancellation conditions have the form of a linear mapping $A
  v=0$, where here, $0 \equiv (0,0,0,0)^T \in {\mathbb R}^4$.  With the basis
  of ${\mathbb R}^6$ ordered as indicated above, $A$ has the matrix form
\beq
A = \left(\begin{array}{cccccc}
  2 & -1 & -1 & 0 & 0 &  0  \\
  3 &  0 &  0 & 1 & 0 &  0  \\
  1 & -8 & -2 & 3 & 0 & -6  \\
  6 & -3 & -3 & 2 &-1 & -1 \end{array}\right) \ .
\label{amatrix}
\eeq
In general, given a linear mapping $A: V \to W$, with ${\rm dim}(V) >
{\rm dim}(W)$, where $V$ and $W$ are two linear vector spaces, if
${\rm rank}(A)$ is maximal, then the kernel (nullspace) of $A$ is
spanned by vectors in a space of dimension ${\rm dim}({\rm
  ker}(A))={\rm dim}(V)-{\rm dim}(W)$. In the present case, $A$ has
maximal rank (equal to 4), so its kernel has dimension 2.  That is,
the most general vector $v \in {\mathbb R}^6$ that is a solution to
$Av=0$ is determined by two independent variables.  Since this is a
linear map, these can, with no loss of generality, be restricted to be
rational, i.e., ${\rm ker}(A)$ in ${\mathbb Q}^6$ is determined by two
independent rational variables. This restriction to rational values is
motivated in order to allow for the embedding of U(1)$_z$ (as well as
U(1)$_Y$ and the rest of $G_{\rm SM}$ ) in a single non-Abelian gauge
symmetry group in the UV, since this embedding would yield rational
values of the U(1)$_z$ charges (as well as rational values of the weak
hypercharges).  Such an embedding would also have the appeal of
avoiding possible Landau singularities in either or both the U(1)$_Y$
and U(1)$_z$ gauge interactions.  As in \cite{adh}, we choose these to
be $z_Q$ and $z_u$. In terms of these charges, the U(1)$_z$ charges of
the other fermions are given by
\beq
z_d = 2z_Q - z_u
\label{zdsol}
\eeq
\beq
z_L = -3z_Q
\label{zlsol}
\eeq
\beq
z_\ell = -(2z_Q + z_u)
\label{zesol}
\eeq
and
\beq
z_\nu = -4z_Q + z_u \ .
\label{znu}
\eeq
The first of these relations can also be written in a form relating the sum
of the U(1)$_z$ charges of the two SU(2)$_L$-singlet quark fields to the
U(1)$_z$ charge of the SU(2)$_L$-doublet quarks, namely
\beq
z_u + z_d = 2z_Q \ .
\label{zudq}
\eeq
The lepton U(1)$_z$ charge assignments imply the analogous relation,
\beq
z_\nu + z_\ell = 2z_L \ .
\label{znueell}
\eeq
With the U(1)$_z$ charge assignments (\ref{zdsol})-(\ref{znu}), both
the quadratic ACC, Eq. (\ref{anom_u1yu1zu1z}), and the cubic ACC,
Eq. (\ref{anom_u1zu1zu1z}), are automatically satisfied. Thus,
although the six ACCs involve two nonlinear equations, the solution is
completely determined by the linear subset of four ACCs.  Thus, 
the U(1)$_z$ charge assignments of the SM fermions that satisfy these six
ACCs are determined by any two, which can be taken as $z_Q$ and $z_u$.
These will be further constrained below.

It is also necessary to ensure that the Yukawa terms for the SM fermions
and the Yukawa term that produces Dirac neutrino masses are invariant
under the gauge group $G$, and, in particular, the U(1)$_z$ factor group in
$G$. We list each of these types of terms (suppressing SU(3)$_c$, SU(2)$_L$,
and generational indices) and the corresponding necessary
and sufficient condition on the $z$ charge of the SM Higgs field, $H$, below:
\beq
{\cal L}_{\rm eff} \supset \bar 
Q_L d_R H + h.c.  \ \Rightarrow \ -z_Q + z_d + z_H = 0
\label{yukd}
\eeq
\beq
{\cal L}_{\rm eff} \supset 
\bar Q_L u_R H^\dagger + h.c. \ \Rightarrow \ -z_Q + z_u - z_H = 0
\label{yuku}
\eeq
\beq
{\cal L}_{\rm eff} \supset 
\bar L_L \ell_R H + h.c. \ \Rightarrow \ -z_L + z_\ell + z_H = 0
\label{yuke}
\eeq
\beq
{\cal L}_{\rm eff} \supset 
\bar L_L \nu_R H^\dagger + h.c. \ \Rightarrow \ -z_L + z_\nu - z_H = 0. 
\label{yuknu}
\eeq
All four of these equations are satisfied with the following U(1)$_z$
charge assignment for the SM Higgs field $H$:
\beq
z_H = -z_Q + z_u \ .
\label{zhiggs}
\eeq

There are several possibilities for neutrino mass terms in this type of model,
depending on U(1)$_z$ charge assignments. If $z_\nu=0$, then the Lagrangian can
contain bare Majorana mass terms of the form (with the generational indices
$a,b$ indicated)
\beq
 \sum_{a,b=1}^3 M^{(R)}_{ab}\nu_{a,R}^T C \nu_{b,R} + h.c. 
\label{nurmasses}
\eeq
(where here $C$ is the Dirac charge conjugation matrix) that are
invariant under the U(1)$_z$-gauge symmetry.  In turn, these can form
the basis for a seesaw mechanism that can naturally explain the small
masses of the observed neutrinos.  If, on the other hand, $z_\nu \ne
0$, then, in order to obtain the Majorana mass terms in
Eq. (\ref{nurmasses}), one would start with terms of the form
\beq
\Big [ \sum_{a,b=1}^3 c^{(R)}_{ab}\nu_{a,R}^T C \nu_{b,R} \Big ] \omega + h.c. 
\label{nuyuk}
\eeq
Substituting the $\omega$ VEV, $\langle \omega \rangle_0 =
v_\omega/\sqrt{2}$ yields the Majorana mass terms of the form
(\ref{nurmasses}) with
\beq
M^{(R)}_{ab} = c^{(R)}_{ab} \frac{v_\omega}{\sqrt{2}} \ . 
\label{nurmasses_yuk}
\eeq
In order for the terms (\ref{nuyuk}) to be invariant under the
U(1)$_z$ gauge symmetry, the U(1)$_z$ charges $z_\nu$ and $z_\omega$
must satisfy the condition $2z_\nu + z_\omega=0$.  With $z_\omega$
taken to be 1, as above (by the normalization of $g_z$), this is
the condition that $z_\nu=-1/2$.  This leads to two classes
\beq
{\rm Class} \ C_1: \ z_\nu=0
\label{znu_class1}
\eeq
and
\beq
{\rm Class} \ C_2: \ z_\nu=-\frac{1}{2} \ .
\label{zu_class2}
\eeq
Combining these conditions with Eq. (\ref{znu}), these two classes of models
are characterized by the respective conditions 
\beq
C_1: \ z_u = 4z_Q 
\label{c1condition}
\eeq
and
\beq
C_2: \ -\frac{1}{2} = -4z_Q + z_u \ . 
\label{c2condition}
\eeq
Thus, in each of these classes of models, the U(1)$_z$ charges of the SM
fermions depend on one input value, which could be taken to be $z_Q$. 

A word is in order concerning possible higher-dimensional operators
contributing to neutrino
masses. Let us define the SU(2)$_L$ tensor \beq (I_{ss})_{ijkm} \equiv
(\epsilon_{ik}\epsilon_{jm} + \epsilon_{im}\epsilon_{jk}) \ .
\label{iss}
\eeq
In the 4D Lagrangian ${\cal L}_{\rm eff}$, the dimension-5 operator 
\beq
\frac{1}{\Lambda_{LLHH}} 
(I_s)_{ijkm} \Big [\sum_{a,b}^3 c^{(L)}_{ab} L^{i \ T}_{a,L} C 
L^j_{b,L}\Big ] H^k H^m + h.c. 
\label{LLHH}
\eeq
involves symmetric combinations of the two SU(2)$_L$ lepton doublets to form an
SU(2)$_L$ isovector, and, similarly, a symmetric combination of the two
SU(2)$_L$ Higgs doublets to form an SU(2)$_L$ isovector, with
the contraction of these two
isovectors to form an SU(2)$_L$ singlet (which is also invariant under
U(1)$_Y$). In Eq. (\ref{LLHH}), $\Lambda_{LLHH}$ represents a relevant mass
scale and the $c^{(L)}_{ab}$ are dimensionless constants. If this operator
occurs, then, via the VEV of the SM Higgs $H$, it yields the Majorana mass
terms $\sum_{a,b=1}^3 (c^{(L)}_{ab}v^2/2)[\nu^T_{a,L} C \nu_{b,L}]$.  This
operator (\ref{LLHH}) has U(1)$_z$ charge $2(z_L + z_H)=2(-4z_Q+z_u)$. Hence,
it is present in Class $C_1$, but would violate the U(1)$_z$ gauge symmetry in
Class $C_2$.  One could also consider other higher-dimension operators that
could contribute to neutrino mass terms.

Since we shall make use of the solution for fermion wave function
centers in the extra dimensions derived in \cite{nuled}, we shall focus
on the version of the model embodied in Class $C_1$ here, where bare
Majorana mass terms for the $\nu_{a,R}$ fields are allowed.
For further details
concerning the procedure for choosing the wave function centers of the
SM fermions, the reader is referred to Refs. \cite{bvd,nuled}.

As noted, owing to Eqs. (\ref{znu_class1}) and (\ref{c1condition}),
this model is completely specified by the values of $z_Q$ and $z_\chi$
(the latter of which is not constrained by anomaly cancellation, since
$\chi$ is a SM-singlet and has a vectorial U(1)$_z$ gauge).  In Table
\ref{z_table} we list the fermion and SM Higgs U(1)$_z$ charges in
this Class 1 version of the theory, in terms of $z_Q$ and,
alternatively, $z_L$, as independent variables.  We restrict these to
values $\lesssim O(1)$ so that the U(1)$_z$ gauge interaction is
perturbative.  We also require that $g_z v_\omega \gg z_H v$ so that
the mixing of the U(1)$_z$ gauge field $C_\mu$ with the neutral
SM gauge fields $A^3_\mu$ and $B_\mu$ is negligibly small, as discussed
in the Appendix. 

\begin{table}
  \caption{\footnotesize{Values of the U(1)$_z$ charges of fermion and
      Higgs fields in the Class $C_1$ version of the model considered
      here. These are expressed in terms of $z_Q$ as the independent
      variable and, equivalently, in terms of $z_L$ as the independent
      variable.}}
\begin{center}
\begin{tabular}{|c|c|c|} \hline\hline
field & $z$ & equiv. \ $z$ 
\\ \hline
$Q_L$      & $z_Q$    & $-(1/3)z_L$  \\
$z_u$      & $4z_Q$   & $-(4/3)z_L$  \\
$z_d$      & $-2z_Q$  & $(2/3)z_L$   \\
$z_L$      & $-3z_Q$  & $z_L$        \\
$z_\nu$    & 0        & 0            \\
$z_\ell$   & $-6z_Q$  & $2z_L$       \\
$z_H$      & $3z_Q$   & $-z_L$       \\
$z_\omega$ & 1        & 1            \\ 
\hline\hline
\end{tabular}
\end{center}
\label{z_table}
\end{table}
%


\section{Further Aspects of the Model}
\label{further_aspects_section}

\subsection{Placement of $\chi_L$ and $\chi_R$ Wave Function Centers in
  the Extra Dimensions}

To proceed, we specify the wave function centers for the fermions in
the extra dimensions. As mentioned above, we shall rely on the choice
of fermion wave function centers given in \cite{nuled} for the
fermions in Eq. (\ref{fermions}). A basic property of the previous
solution was the necessity to separate the wave function centers of
the quarks in the extra dimensions from those of the leptons, in order
to suppress contributions to proton and baryon-number-violating 
bound neutron decays. This is evident in Fig.  1 of Ref. \cite{nuled}.
To complete the specification of fermion wave function centers, we
thus must make choices for the wave function centers of the $\chi_L$
and $\chi_R$ fields in the extra dimensions. These choices depend on
(i) the mass $m_{\chi}$, which, in turn, is determined, for a given
choice of $m_{\chi,4+n}$, by $\|\eta_{\chi_L} - \eta_{\chi_R} \|$; and
(ii) the distances $\|\eta_{\chi_L} - \eta_{f_{a,k'}} \|$ and
$\|\eta_{\chi_R} - \eta_{f_{a,k'}} \|$, where $1 \le a \le 3$ and
$k'=L,R$ for SM fermions $f_{a,k'}$.
We use an iterative procedure to find acceptable values for these
quantities.

There are actually several different types
of dark matter models that we can construct.  The two types that we
will focus on here involve dark matter that is initially in thermal
equilibrium with SM fields at high temperature in the early universe
and freezes out as the temperature decreases below a certain value
denoted $T_{\rm f.o.}$. These are (i) leptophilic DM, if the $\chi_L$
and/or $\chi_R$ wave function centers are closer to those of the
leptons; (ii) hadrophilic DM, if the $\chi_L$ and/or $\chi_R$ wave
function centers are closer to those of the quarks. In a thermal dark
matter framework, the freeze-out temperature is given in terms of the
dark matter particle mass by the approximate relation 
$k_BT_{\rm f.o.}/m_\chi \simeq 0.05$ (e.g. \cite{battaglieri} and
references therein).  A third type of dark matter model is obtained if we
choose the $\chi_L$ and $\chi_R$ wave function centers to be far from
the wave function centers of both the quark and SM fermions.  In this
case, the exponential suppression of the interactions of the
dark matter $\chi$ fermions with SM fermions may be so severe that the dark
matter is not in thermal equilibrium with SM fields during the
relevant time period in the early universe.  (This also depends on the
degree of suppression of the mixing of the $C_\mu$ gauge fields with
$A^3_\mu$ and $B_\mu$ gauge fields.)  In this version of the model, rather
than decreasing with time, the dark matter relic density is initially
negligible and gradually builds up, eventually freezing in at the observed
value \cite{hall_freeze_in,uv_freeze_in,freeze_in_review}.  In the present
work we focus on the first two versions of our model; the
freeze-in version merits future study.

In Fig. \ref{fermion_centers_chi_near_ell_figure} we show an illustrative
choice of $\chi_L$ and $\chi_R$ wave function centers in the $n=2$ extra
dimensions for the leptophilic version of the model.
In Fig. \ref{fermion_centers_chi_near_q_figure} 
we show an illustrative choice
of $\chi_L$ and $\chi_R$ wave function centers for the hadrophilic version of
the model.  


\begin{figure}
  \begin{center}
\includegraphics[height=0.9\linewidth]{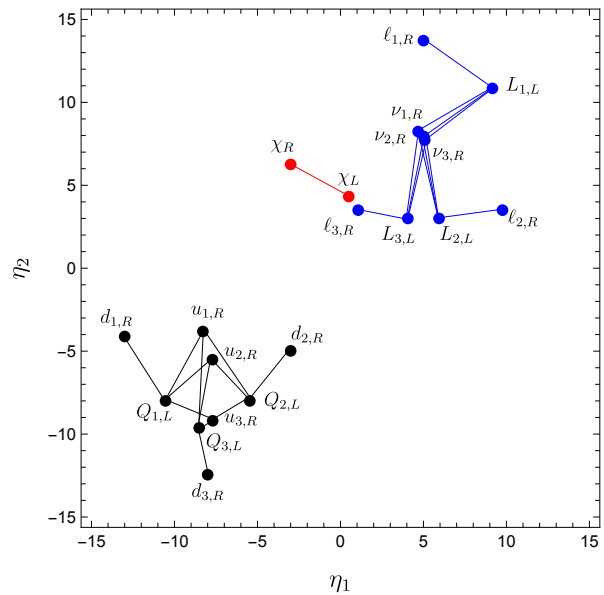}
  \end{center}
  \caption{Plot showing locations of fermion wave function centers in
    the leptophilic version of the model. As defined in the text, the
    numerical subscript on each fermion field is the generation index.
    Toroidal compactification is used, so that
    $\eta_\lambda$ is equivalent to $\eta_\lambda \pm \mu L = \eta_L
    \pm 30$. In the online figure, the locations of the
    lepton wave functions are indicated with the following colors:
    leptons in blue, $\eta_{\chi_{L,R}}$ in red, and quarks in black.}
    \label{fermion_centers_chi_near_ell_figure}
\end{figure}


\begin{figure}
  \begin{center}
\includegraphics[height=0.9\linewidth]{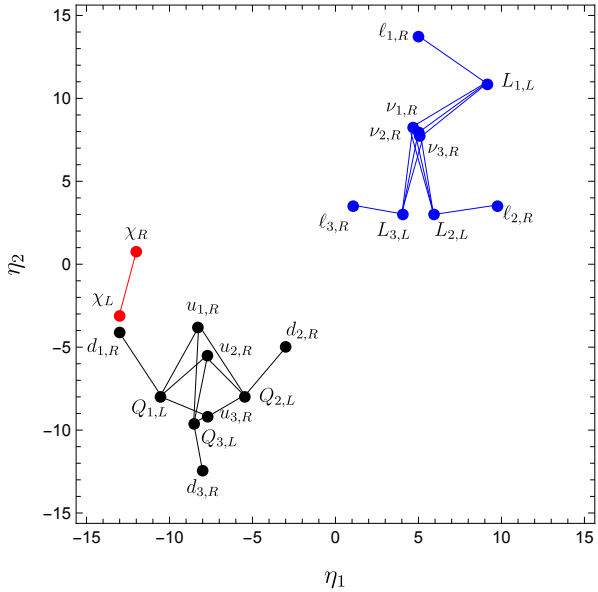}
  \end{center}
  \caption{Plot showing locations of fermion wave function centers in
    the hadrophilic version of the model.  The notation is as in the
    previous figure.}
\label{fermion_centers_chi_near_q_figure}
\end{figure}


\subsection{Stability of $\chi$}

A necessary condition for a dark matter particle is that it must live
longer than the age of the universe, $t_U$.  According to the
consensus cosmological model, $t_U = 13.8 \times 10^9 = 4.35 \times
10^{17}$ sec \cite{pdg_dm}.  However, this condition is not, general,
sufficient; the dark matter must also be sufficiently long-lived to
obey constraints from indirect-detection searches.  For example, if
tree-level and/or loop-level decays yield photons, then constraints
from indirect-detection searches, such as those of Fermi-LAT, imply
that the lifetime of a dark matter particle $\tau_{DM}$ should obey
the lower bound $\tau_{DM} \gsim 10^{25}$ sec \cite{baring}. In order
to satisfy this constraint in the simplest way, we shall assume that
the theory is invariant under a ${\mathbb Z}_2$ symmetry that operates
only on $\chi$ and with respect to which $\chi$ is odd.  This
guarantees that $\chi$ is stable.


\section{Calculation of Relic Abundance of Dark Matter}
\label{relic_abundance_section}

In order to account for the observed relic density of dark matter with
a non-self-conjugate DM particle such as the Dirac $\chi$ fermion that
we use, it is necessary that the thermal average of the $\chi
\bar\chi$ annihilation cross section multiplied by (relative) velocity,
$v_r$, should satisfy the following relation applicable for the time of
freeze-out, for the mass range $10 \ {\rm GeV} \ \lsim m_\chi \lsim 100$
TeV \cite{kolb_turner}-\cite{tulin_yu},\cite{steigman_dm,steig2}
\beq
\langle \sigma v_r \rangle \simeq 4 \times 10^{-26} \ \frac{{\rm cm}^3}{s} \ .
\label{sigmav_needed}
\eeq
As will be shown below, this mass range includes the preferred range
of values of $m_\chi$ in our model.  Since our dark matter particle is
not non-self-conjugate, there is also the related question of a
possible nonzero net number asymmetry $N_\chi - N_{\bar\chi}$,
analogous to the baryon asymmetry in the universe. While asymmetric
dark matter models are of interest \cite{zurek_adm}, we do not assume
a $N_\chi - N_{\bar\chi}$ number asymmetry here.

We will calculate the value of $\langle \sigma v_r \rangle$ in our
model and set it equal to the value in Eq. (\ref{sigmav_needed}) to
constrain the model.  For this purpose, we first determine the
dominant contribution to the reaction in which $\chi$ and $\bar\chi$
annihilate, yielding SM particles.  Given the suppression in the
mixing of the U(1)$_z$ gauge field $C_\mu$ with the SU(2)$_L$
$A^3_\mu$ and the U(1)$_Y$ $B_\mu$ gauge fields, we may obtain an
approximate estimate of the relic abundance by calculating the
contribution from a process in which the $\chi$ and $\bar\chi$
annihilate to produce a virtual U(1)$_z$ gauge boson $C$ in the $s$
channel, which then materializes into the final-state SM
fermion-antifermion pair, $f \bar f$.  Recall that with the
spontaneous breaking of the U(1)$_z$ gauge symmetry, the U(1)$_z$
vector boson, $C$, picks up a mass $m_C = g_z v_\omega/2$ (where, as
noted above, the mixing with $A^3$ and $B$ is negligibly small, so
that $C$ is both an interaction eigenstate and, to a very good
approximation, a mass eigenstate).  Our solution to fit the relic
abundance of the dark matter will entail the inequality $m_C \gg
m_\chi$, so that the the momentum dependence in the U(1)$_z$ vector
propagator is negligible. Hence, denoting $s=(p_\chi +
p_{\bar\chi})^2$ as the center-of-mass (CM) energy squared in the
$\chi \bar\chi$ annihilation reaction at relevant times in the early
universe, where $p_\chi$ and $p_{\bar\chi}$ are the four-vectors of
the $\chi$ and $\bar\chi$, we have
\beq
\frac{g_z^2}{m_C^2 - s} \simeq \frac{g_z^2}{m_C^2} = \frac{4}{v_\omega^2} \ . 
\label{cpropagator}
\eeq
In the $d=4+n$ space, the resultant amplitude (suppressing the generational
index $a$ on the SM fermions $f_{a,L}$ and $f_{a,R}$) is proportional to 
\beqs
[\bar\chi \gamma_\mu \chi] [\bar f \gamma^\mu f] &=&
\Big ( 
[\bar \chi_L \gamma_\mu \chi_L] + ([\bar \chi_R \gamma_\mu \chi_R] \Big ) \, 
\times \cr\cr
&\times& \Big (
[\bar f_L \gamma_\mu f_L] + [\bar f_R \gamma_\mu f_R] \Big ) \ . 
\label{chichiff}
\eeqs
Together with the $\gamma$ matrix structure and Dirac spinors, these operator
products involve a sum of products of $y$-dependent wave functions, namely
\begin{widetext}
\beqs
A_f^2 \Big [ 
e^{-2( \| \eta-\eta_{\chi_L} \|^2 + \| \eta-\eta_{f_L} \|^2)} +  
e^{-2( \| \eta-\eta_{\chi_L} \|^2 + \| \eta-\eta_{f_R} \|^2)} +
e^{-2( \| \eta-\eta_{\chi_R} \|^2 + \| \eta-\eta_{f_L} \|^2)} +
e^{-2( \| \eta-\eta_{\chi_R} \|^2 + \| \eta-\eta_{f_R} \|^2)} \Big ] \ .
\label{eefactors}
\eeqs
\end{widetext}
Upon integration over the $y$ coordinates, the first of these yields
a term proportional to $e^{-\|\eta_{\chi_L}-\eta_{f_L} \|^2}$, and so forth 
for the others (see the general integration formula (A2) in \cite{bvd}). It
follows that the dominant contribution to the low-energy effective Lagrangian
from these four-fermion operators in $d=6$ dimensions arises from the term with
the smallest distance between (chiral components of) $\chi$ and $f$. For
the case where this smallest-distance criterion picks out the four-fermion
product $[\bar \chi_k \gamma_\mu \chi_k][\bar f_{a,k'}\gamma^\mu f_{a,k'}]$
with generational index $a$ for the SM fermion $f_{a,k'}$ \cite{gennote}
and chiralities $(k,k')$ taking value(s) in the set
$\{(L,L), \ (L,R), \ (R,L), \ (R,R) \ \}$,  the dominant
term is then proportional to 
\beq
[\bar \chi_k \gamma_\mu \chi_k][\bar f_{a,k'} \gamma^\mu f_{a,k'}] \ .
\label{4fermion_operator}
\eeq
Carrying out the integration over the wave functions in the extra
dimensions, as in \cite{nnb02,bvd,nuled}, we obtain the following term
in the low-energy 4D effective Lagrangian:

\beq
    {\cal L}^{(\chi_k f_{a,k'})}_{\rm eff, \ int.} =
    \frac{c^{(\chi_k f_{a,k'})}}
    {\Lambda_{\rm eff}^2}
[\bar\chi_k\gamma_\mu \chi_k][\bar f_{a,k'}\gamma^\mu f_{a,k'}] + h.c.
\label{leff_chi_f}
\eeq
where
\beq
c^{(\chi_k f_{a,k'})} = \frac{\xi^2}{\pi} \,
e^{-\|\eta_{\chi_k}-\eta_{f_{a,k'}} \|^2 } 
\label{c_chi_f}
\eeq
and
\beq
\frac{1}{\Lambda_{\rm eff}^2} = \frac{g_z^2 z_\chi z_{f_{k'}}}{m_C^2} = 
\frac{4z_\chi z_{f_{k'}}}{v_\omega^2} \ .
\label{lamda_factor}
\eeq
%


\subsection{Leptophilic Dark Matter}

We proceed to calculate the relic abundance of the dark matter
in the various versions of the model, beginning with the leptophilic version. 
With our illustrative placement of the $\chi_L$ and $\chi_R$ wave function 
centers shown in Fig. \ref{fermion_centers_chi_near_ell_figure}, the
minimal-distance wave function pair links $\chi_k = \chi_L$ with
$f_{a,k'}=\ell_{3,R} \equiv \tau_R$.  The resultant cross section is
\begin{widetext}
\beq
\sigma = \frac{s}{48\pi\Lambda_{\rm eff}^4} \, |c^{(\chi_L \tau_R)}|^2 \, 
  \bigg [ \frac{1-\frac{4m_\tau^2}{s}}
    {1-\frac{4m_\chi^2}{s}} \bigg ]^{1/2} \,
  \bigg [1-\frac{(m_\chi^2 + m_\tau^2)}{s} + \frac{4m_\chi^2 m_\tau^2}{s^2}
    \bigg ] \ ,
  \label{sigma_chi_tau}
  \eeq
\end{widetext}
where
\beq
c^{(\chi_L \tau_R)} = \frac{\xi^2}{\pi}
e^{-\|\eta_{\chi_L} - \eta_{\tau_R} \|^2} \ .
\label{c_chi_L_tau_R}
\eeq
In the center-of-mass frame, the magnitudes of the velocities of the
colliding $\chi$ and $\bar \chi$ are given by $v_{\chi,CM} =
v_{\bar\chi,CM}= \sqrt{1-(4m_\chi^2/s)}$, and their relative velocity is
$v_r = 2v_{\chi,CM}$.  The quantity that enters into the determination
of the relic density is $\langle \sigma v_r \rangle$ in general and,
in particular, at the time of freeze-out.  Since this freeze-out
occurs when the temperature satisfies $k_BT_{f.o.}/m_\chi \simeq 0.05$
\cite{swo,gondolo_gelmini}, it follows that the $\chi$ fermions are
moderately nonrelativistic at this time, and $s \simeq 4m_\chi^2$. As
the temperature decreases to $T_{\rm f.o.}$, just before the dark
matter $\chi$ fermions drop out of thermal equilibrium, and using the
nonrelativistic equipartition theorem, we have \beq \frac{m_\chi
  v_{\chi,CM}^2}{2} = \frac{3}{2}k_BT \sim \frac{3}{2} \Big (
\frac{m_\chi}{20} \Big ) \ ,
\label{vfo}
\eeq
so $v_{\chi,CM}^2 \simeq 0.08 \ll 1$  For approximate estimates, this
motivates an expansion of $\langle \sigma v_r \rangle$ in powers of
$v_{\chi,CM}$. Performing this expansion, using the fact that
$v_{\chi,CM}^2 \simeq 0.08 \ll 1$, and anticipating the 
result that $m_\tau/m_\chi \ll 1$ from our fit to the relic density,
we thus obtain the approximate analytic formula
\beqs
\langle \sigma v_r \rangle &=& \frac{|c^{(\chi_L\tau_R)}|^2 m_\chi^2}
        {8\pi \Lambda_{\rm eff}^4} \, 
        \Big [ 1 + O(v_{\chi,CM}^2) \Big ] \cr\cr
        & \simeq & \frac{|c^{(\chi_L\tau_R)}|^2 m_\chi^2}
        {8\pi \Lambda_{\rm eff}^4} \ . 
\label{sigmav_result}
\eeqs
Setting this expression for $\langle \sigma v_r \rangle_{\rm f.o.}$
equal to the value $4 \times 10^{-26}$ \ cm$^3$/s in
Eq. (\ref{sigmav_needed}), inserting Eq. (\ref{c_chi_L_tau_R}) for
$c^{(\chi_L \tau_R)}$, and substituting $\xi=30$, we obtain
\beq
2\ln \Big ( \frac{\Lambda_{\rm eff}}{100 \ {\rm TeV}} \Big )
-\ln \Big ( \frac{m_\chi}{10 \ {\rm TeV}} \Big ) 
+\|\eta_{\chi_L} - \eta_{\tau_R} \|^2 = -0.02 .
\label{dmcondition}
\eeq
This, then, is the condition for the parameters $\Lambda_{\rm eff}$,
$m_\chi$, and $\|\eta_{\chi_L} - \eta_{\tau_R} \|^2$ in order for our
model to produce the observed dark matter relic density.  It is easy
to choose parameters that satisfy this condition. For example, we may
take $\|\eta_{\chi_L} -
\eta_{\chi_R} \|=4$, which gives $m_\chi=10$ TeV, and
$\|\eta_{\chi_L} - \eta_{\tau_R} \|=1$, as shown in Fig.
\ref{fermion_centers_chi_near_ell_figure}.  Combined with
$\Lambda_{\rm eff}=60$ TeV, this satisfies Eq. (\ref{dmcondition}).

To confirm the accuracy of this approximate analytic approach to the
calculation of the relic dark matter density, we have carried out a
full numerical computation of the relic density and direct detection
cross-section using {\tt micrOMEGAs} \cite{micromegas} (which computes
the matrix elements using {\tt CalcHEP} \cite{calchep} with {\tt
  LanHEP} inputs \cite{lanhep}). As is evident in
Figs. \ref{relic_leptophilic_analytical} and \ref{relic_leptophilic},
we find excellent agreement between our approximate analytic results
and the results of the full numerical calculation.  In general, as
shown by Eq. (\ref{dmcondition}), Figs. \ref{relic_leptophilic_analytical},
and \ref{relic_leptophilic}, one can easily choose the parameters
$\Lambda_{\rm eff}$,
$m_\chi$ (and hence $\|\eta_{\chi_L}-\eta_{\chi_R}\|$ via Eq. (\ref{mchi4d})),
and $\|\eta_{\chi_L}-\eta_{\ell_{3,R}}\|$ to obtain the observed relic
dark matter density.  A word is in order concerning the upper bound on thermal
WIMP dark matter particles from unitarity
\cite{gkbound}-\cite{smirnov_beacom}. After refinements, the current value
of this upper bound is $m_\chi < 100$ TeV for non-self-conjugate dark matter
\cite{smirnov_beacom}, and this is the reason that we have shown fits
with $m_\chi$ bounded above by this limit. The range of $m_\chi$ values for
which we find satisfactory fits to the relic abundance of dark matter is
thus consistent with the unitarity upper bound.

It should be emphasized that there is considerable freedom in the
model concerning the placement of the wave function centers for
$\chi_L$ and $\chi_R$ relative to the SM leptons in the extra
dimensions.  For example, we could have chosen the $\chi$ wave
function centers to have $\chi_L$ and $\chi_R$ reversed, so that it is
$\chi_R$ that is closer to the wave function center for
$\ell_{3,R}$. In this case, the dominant four-fermion operator
relevant for $\chi \bar \chi$ annihilation to SM fermions would have
been given by Eq. (\ref{4fermion_operator}) with $f_{a,k'}=\ell_{3,R}$
but with $k=R$ instead of $k=L$.  However, this would lead to the same
expression (\ref{sigma_chi_tau}) for $\langle \sigma v_r
\rangle$. Moreover, rather than placing the wave function center for
$\chi_L$ or $\chi_R$ closer to $\ell_{3,R}$, we could, instead, have
placed it closer to a different lepton such as $\ell_{2,R}$ (staying
within the leptophilic version of the model).  This would not change
our general conclusions concerning the ability of the model to
successfully account for the relic density of dark matter. 

\begin{figure}
\begin{center}
\includegraphics[height=0.9\linewidth]{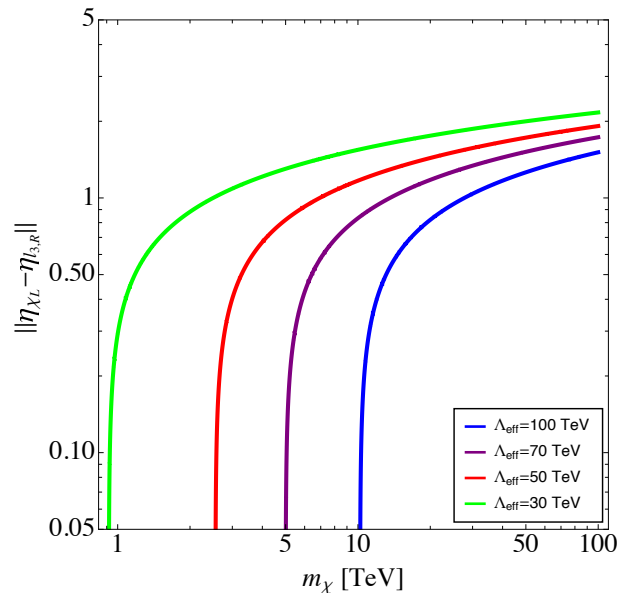}
  \end{center}
  \caption{Values of $m_\chi$ and $d_{\chi \tau} =
    \|\eta_{\chi_L}-\eta_{\tau_R} \|$ that yield the observed dark matter
    relic density in the leptophilic
    version of the model, as given by the approximate analytic solution in
    eq. (\ref{dmcondition}). Curves are plotted for several values of
  $\Lambda_{\rm eff}$.}
\label{relic_leptophilic_analytical}
\end{figure}

\begin{figure}
\begin{center}
\includegraphics[height=0.9\linewidth]{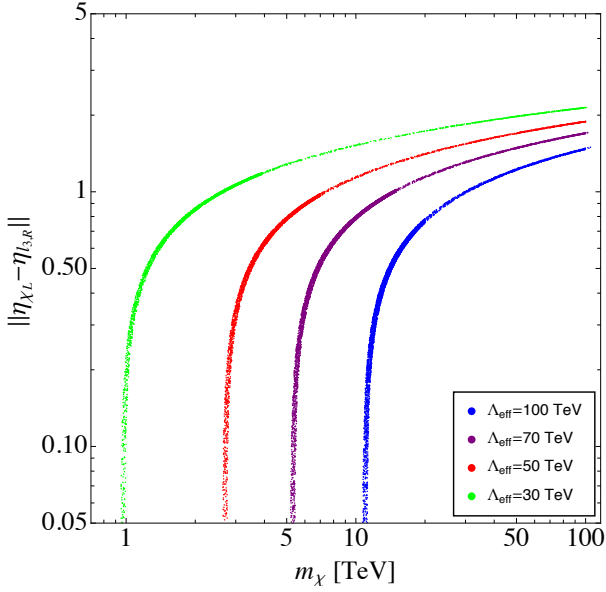}
  \end{center}
  \caption{Values of $m_\chi$ and $d_{\chi \tau} =
    \|\eta_{\chi_L}-\eta_{\tau_R} \|$ that yield the observed dark matter
    relic density in the leptophilic
    version of the model, as calculated numerically using {\tt micrOMEGAs}.
  Curves are plotted for several values of $\Lambda_{\rm eff}$.}
\label{relic_leptophilic}
\end{figure}

\subsection{Hadrophilic Dark Matter}

As indicated in Fig. \ref{fermion_centers_chi_near_q_figure}, one can
also choose the $\chi_L$ and $\chi_R$ wave function centers to lie
closer to the quarks. Hence, in this case the exponential suppression
effect in the low-energy, long-distance effective field theory
resulting from the large separation of the dark matter from the lepton
wave function centers in the extra dimensions means that the dominant
interaction of the dark matter is with quarks.  This version of the model
may thus be called hadrophilic (and also leptophobic). 

Direct detection searches \cite{xenon1T, LUX, PandaX, PICO} have
yielded stringent upper limits on the spin-independent and
spin-dependent cross sections of the dark matter scattering with
nucleons. As the nucleons interacting with the dark matter are
primarily composed of first-generation quarks, $u$ and $d$, these
direct-detection upper bounds are relevant only if the $\chi_L$ and/or
$\chi_R$ wave function centers are located near to a first-generation
quark. As in the case of leptophilic dark matter, one can achieve the
observed relic abundance while simultaneously satisfying the
direct-detection bounds by localizing the dark matter near to
higher-generation quarks in the extra dimensions.  Hence, the only
nontrivial case to consider is the case in which the dark matterr is
localized in the extra dimensions near to a first-generation quark,
taken here to be $d_{1,R}$ for illustrative purposes. Furthermore, for
definiteness, let us assume that $\chi_L$ is located closer to
$d_{1,R}$. The same analysis would apply if $\chi_R$ were closer to
$d_{1,R}$. The constraints on the parameters of the model would be
similar for the cases when the dark matter is not localized near a
first-generation quark, the only difference being that the bounds from
direct-detection searches would not yield significant constraints.

Let us proceed to calculate the condition that yields the observed
relic abundance. The cross-section for $\bar\chi \chi \to \bar d_{1}
d_{1}$ is similar to Eq. (\ref{sigma_chi_tau}), with the inclusion of
an additional color factor of $N_c=3$. Therefore, the cross section is
  \beqs
 \sigma &=& \frac{s}{16 \pi\Lambda_{\rm eff}^4} \,  |c^{(\chi_L d_{1,R})}|^2 \,
 \bigg [ \frac{1-\frac{4m_d^2}{s}} {1-\frac{4m_\chi^2}{s}} \bigg ]^{1/2}
 \times \cr\cr
 &\times&
 \bigg [1-\frac{(m_\chi^2 + m_d^2)}{s}+\frac{4m_\chi^2 m_d^2}{s^2} \bigg ]
 \ ,
 \label{sigma_chi_d}
 \eeqs
where 
\beq
c^{(\chi_L d_{1,R})}=\frac{\xi^2}{\pi}
e^{-\|\eta_{\chi_L}-\eta_{d_{1,R}}\|^2} \ .
\label{c_chi_L_d_R}
\eeq
(In Eq. (\ref{sigma_chi_d}) we retain the $m_d$ terms only for
symmetry; they are negligibly small.)  From Eq. (\ref{lamda_factor}),
it is evident that $\Lambda_{\rm eff}$ depends on $z_{d_{1,R}}$. To
simplify the notation, we absorb this dependence in the notation and
define $\Lambda_{\rm eff}^{(\chi_L d_{1,R})} \equiv \Lambda_{\rm
  eff}$.  As before, we expand $\langle \sigma v_r \rangle$ in powers
of $v_{\chi, CM}$, and obtain
\beqs
\langle \sigma v_r \rangle &=& \frac{3 m_\chi^2}
  {8 \pi \Lambda_{\rm eff}^4} |c^{(\chi_L d_{1,R})}|^2
  \bigg[ 1 + \mathcal{O}(v_{\chi, CM}^2) \bigg] \cr\cr
  &\simeq& \frac{3 m_\chi^2}{8\pi \Lambda_{\rm eff}^4}|c^{(\chi_L d_{1,R})}|^2
  \cr\cr
  &=& \frac{3 m_\chi^2}{8\pi \Lambda_{\rm eff}^4} \,
  \bigg ( \frac{\xi^2}{\pi} \bigg )^2 \,
  e^{-2 \|\eta_{\chi_L}-\eta_{d_{1,R}}\|^2} \ .
\label{sigmav_result_chid}
\eeqs
Equating this with
$\langle \sigma v_r \rangle \simeq 4 \times 10^{-26} {\rm cm}^3/{\rm s}$
and substituting $\xi=30$, we obtain
\beq
2\ln \Big ( \frac{\Lambda_{\rm eff}}{100 \ {\rm TeV}} \Big )
-\ln \Big ( \frac{m_\chi}{10 \ {\rm TeV}} \Big ) 
+\|\eta_{\chi_L} - \eta_{d_{1,R}} \|^2 = 0.53 \ .
\label{dmcondition_chid}
\eeq
For example, we can choose $\|\eta_{\chi_L}-\eta_{d_{1,R}}\|=1$,
$\Lambda_{\rm eff}=79$ TeV, and $\|\eta_{\chi_L}-\eta_{\chi_R}\|=4$ as
shown in Fig. \ref{fermion_centers_chi_near_q_figure}, which yields
$m_\chi=10$ TeV. This illustrative choice satisfies
Eq. (\ref{dmcondition_chid}).

We use the {\tt nucleonAmplitudes} routine in the {\tt micrOMEGAs}
\cite{micromegas} to numerically evaluate the DM-nucleon elastic
scattering amplitude. We then use this to calculate the
spin-independent and spin-dependent cross-sections of the DM-nucleon
scattering for the parameter values satisfying the relic density
constraint.  We employ the Xenon1T upper bounds \cite{xenon1T} on the
DM-nucleon scattering cross section to constrain our model. For our
case, the spin-dependent cross sections are well below the current
limits \cite{xenon1T}, whereas the spin-independent constraints
provide non-trivial bounds on the parameters of the hadrophilic
version of the model.
\begin{figure}
  \begin{center}
\includegraphics[height=0.8\linewidth]{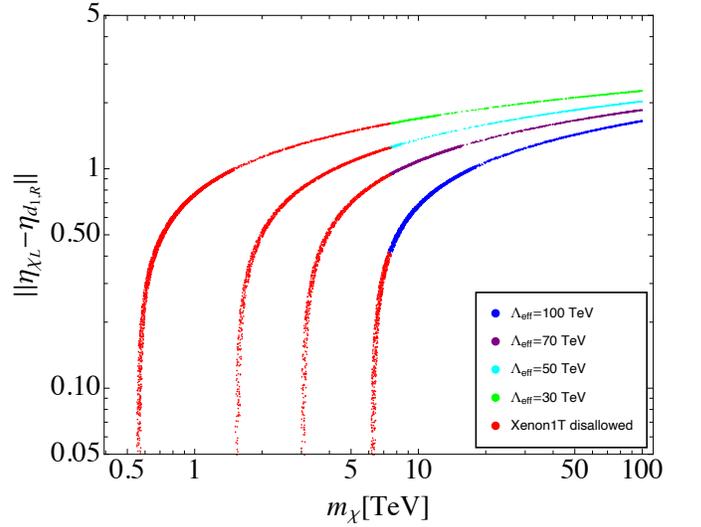}
  \end{center}
  \caption{Plot of values of $m_\chi$ and $d_{\chi d}=\|\eta_{\chi_L}-
    \chi_{d_{1,R}}\|$ yielding dark matter relic density in agreement
    with observation in the hadrophilic version of the model, as
    functions of $\Lambda_{\rm eff}$.
    The red points produce direct-detection
    cross sections that are excluded from the Xenon1T bound \cite{xenon1T},
    while the other points represent parameter values that are allowed by
    both the observed dark matter relic density
    and the direct-detection bound.} 
\label{relic_hadrophilic}
\end{figure}
We show these constraints in Fig. \ref{relic_hadrophilic}. The red
points yield spin-independent DM-nucleon cross-section values that are
excluded from the Xenon1T bound \cite{xenon1T}. The other colored
points are allowed by both the relic abundance constraint and
direct-detection bounds. The agreement of the numerical results, as
shown in Fig. \ref{relic_hadrophilic} for the contours yielding the
observed relic density, with the approximate analytical formula in
Eq. (\ref{dmcondition_chid}) is excellent. A general feature is that as
$\Lambda_{\rm eff}$ decreases, the DM-nucleon scattering cross section
increases, and this, in turn, requires a larger DM-quark separation in the
extra dimensions in order to be consistent with the experimental
bounds. Analogous bounds from LUX \cite{LUX} and PandaX-II \cite{PandaX}
are somewhat less stringent in the $\chi$ mass range relevant here. 
Our general conclusion for the hadrophilic version of our model is thus
similar to the conclusion that we reached for the leptophilic version,
namely that there is a substantial range of choices for the parameters
$\Lambda_{\rm eff}$, $m_\chi$, and $\|\eta_{\chi_L}-\eta_{d_{1,R}}\|$
such that the model yields a value for the dark matter relic
density that matches observation and is also in accord with bounds from
direct-detection searches for dark matter. 


\section{Experimental and Observational Tests}
\label{experimental_tests_section}

In this section we discuss experimental tests of the model.
One can distinguish two types of tests: (i) searches for evidence of
extra spatial dimentions {\it per se}, independent of issues of dark matter,
and (ii) searches for particle dark matter of the type considered here.

Concerning tests of type (i), we again note that the type of
extra-dimensional model used here \cite{as,ms} is quite different from
the type in which there is a low scale of quantum gravity
\cite{dif,ged}, as is clear from the fact that the compactification
scale here is $L = 2 \times 10^{-19}$ cm, much smaller than the
typical micron-size compactification scale in models considered in
\cite{ged}.  As discussed in \cite{nuled,kk}, bounds on
flavor-changing neutral current effects and precision electroweak data
provide significant constraints on this type of extra-dimensional
model.  Owing to the high energy scale $\Lambda_L=100$ TeV
corresponding to the inverse compactification scale, experiments at
the Tevatron and Large Hadron Collider do not yield constraints on the
extra-dimensional aspects of the model.

With regard to searches for dark matter via direct detection, the two
leptophilic and hadrophilic versions of the model have quite
different properties.  As noted above, the leptophilic version of
the model easily satisfies bounds from direct detection searches. 
In contrast, in the hadrophilic version of the model, if one places
the wave function center of $\chi_L$ and/or $\chi_R$ near to a
first-generation quark in the extra dimensions, then the bounds 
from direct detection dark matter searches are non-trivial, as shown in
Fig. \ref{relic_hadrophilic}. Future direct-detection searches will
improve the bounds on spin-independent and spin-dependent DM-nucleon
cross-sections, which will further constrain the  parameters shown in
Fig. \ref{relic_hadrophilic}. 

With regard to indirect detection, since we have constructed the model
so that the $\chi$ dark matter fermion is stable, we only discuss
constraints due to $\chi \bar\chi$ annihilation.  This annihilation
yields SM particles in the final states, and these have been searched
for by several ground-based and space-based instruments. The current
bounds on the thermal-average cross-section from indirect-detection
experiments such as Fermi-LAT and MAGIC 
exclude thermally produced DM particle masses of up to 10-100 GeV
\cite{fermiLAT,fermiLAT_MAGIC}.  As is evident from our results in
Figs. \ref{relic_leptophilic} and 
\ref{relic_hadrophilic}, the range of masses $m_\chi$ for which
we fit the observed cosmological dark matter relic density
is consistent with these bounds from indirect-detection experiments.
Future indirect-detection searches plan to achieve sensitivity to
thermal dark matter cross-sections in the multi-TeV range and
should improve bounds to constrain the parameter
spaces of the thermally produced DM models that we have presented.

As regards collider searches for dark matter particles,
because $m_\chi$ is typically in the multi-TeV range and interacts very weakly
with SM particles, these searches do not probe our model very stringently.
Clearly, collider searches at higher-energies would improve the reach of
these probes.

It is also of interest to investigate the strength of the
self-interactions of the DM $\chi$ fermions. One motivation for this
is that there have been arguments that, although collisionless cold
dark matter, as embodied in the standard $\Lambda$CDM paradigm (with
$\Lambda$ referring to dark energy), fits cosmological data on
large-distance scales of order 100 Mpc to Gpc
\cite{nfw,binney_tremaine}, it may encounter problems on
shorter-distance scales of order 1-20 kpc.  These problems include (i)
the core-cusp problem, i.e., the CDM prediction of too high mass
densities at centers of galaxies; (ii) the CDM prediction of
substantially more dwarf satellite galaxies than are observed
(although, the number of observed dwarf galaxies associated with the
Milky Way has been substantially increased by recent observations); and
(iii) the so-called ``too-big-to-fail'' problem of star formation in
satellite galaxies \cite{moore}-\cite{boylan_kolchin}. These have led
to the consideration of models in which the dark matter has
substantial self-interactions \cite{bullock_boylan_kolchin,tulin_yu},
\cite{spergel_steinhardt}- \cite{kaplinghat2020}. Inclusion of baryon
feedback effects in dark matter simulations may alleviate or remove
these problems \cite{springel}-\cite{peter}. (Other approaches to dark
matter that address these issues have also been studied, e.g.,
\cite{wdm}-\cite{fdm}.)  Further observational work and 
improvement of theoretical modelling of structure formation should
elucidate how serious the possible problems are for the $\Lambda$CDM
paradigm. However, at least this motivates an assessment of the size of
the $\chi$ self-interactions here.

The self-interactions of the dark matter particle in our model 
involve $\chi \bar \chi \to \chi \bar \chi$
reactions and $\chi \chi \to \chi \chi$ (and $\bar \chi \bar \chi \to
\bar \chi \bar \chi$) reactions, which are the DM
analogues of Bhabha and M\o{}ller scattering in quantum electrodynamics.
The $\chi \bar \chi \to \chi \bar \chi$ involves exchange of the U(1)$_z$
vector boson $C$ in the $s$ channel and $t$ channel, while the
$\chi \chi \to \chi \chi$ reaction involves only the $C$ exchange in the
$t$ channel. For our purposes, a rough estimate of these cross sections will
be sufficient.  Since $m_C \gg m_\chi$, the propagator is well approximated
by a constant.  Thus, the amplitudes for the reactions have a prefactor 
$z_\chi^2g_z^2 /m_C^2 = 4z_\chi^2/v_\omega^2$. The operator product in
the $d=6$ dimensional space is the analogue of Eq. (\ref{chichiff})
with $f=\chi$, and involves the products of Gaussian factors listed in
Eq. (\ref{eefactors}) with this substitution $f=\chi$.  Of the four
products of Gaussians, the dominant ones are those in which the
$\chi$ and $\bar\chi$ wave function centers are at the same point in the
extra dimensions, namely 
$A_f^2 e^{-4 \| \eta-\eta_{\chi_L} \|^2}$ and
$A_f^2 e^{-4 \| \eta-\eta_{\chi_R} \|^2}$.  After integration of these
operator products over the extra dimensions, we obtain the resultant
low-energy 4D effective operators for the $\chi$ self-interaction (SI):
\beqs
    && {\cal L}_{\rm SI,4D,eff} \simeq \Big ( \frac{4z_\chi^2}{v_\omega^2}
    \Big ) \, \Big ( \frac{\xi^2}{\pi} \Big ) \times \cr\cr
    &\times&
    \bigg [ [\bar\chi_L \gamma_\mu \chi_L][\bar\chi_L \gamma^\mu \chi_L] +
      [\bar\chi_R \gamma_\mu \chi_R][\bar\chi_R \gamma^\mu \chi_R] \bigg ] \ .
    \cr\cr
    && 
  \label{chi4f}
\eeqs
The resultant cross sections for the $\chi \bar \chi \to \chi \bar\chi$
and $\chi \chi \to \chi \chi$ reactions at a center-of-mass energy
$s \gsim 4m_\chi^2$ are 
\beq
\sigma_{SI} \sim \Big ( \frac{4z_\chi^2}{v_\omega^2}\Big )^2 \,
                    \Big ( \frac{\xi^2}{\pi} \Big )^2  \, m_\chi^2 R_2 \ , 
\label{sigma_4chi}
\eeq
where $R_2$ denotes the two-body final-state phase space, $R_2 <
1/(8\pi)$.  Using the illustrative values $g_z \sim 0.5$, $z_\chi \sim
1$, $v_\omega \simeq 400$ TeV together with a value $m_\chi=10$ TeV in
accord with our fit to the DM relic density, and again using the
$\xi=30$ characterizing our basic extra-dimensional framework, we find
that these cross sections are $\sim 10^{-35}$ cm$^2$ and hence
$\sigma_{SI}/m_\chi \sim 10^{-39} \ {\rm cm}^2/{\rm GeV} \sim 10^{-15}
\ {\rm cm}^2/{\rm g}$ (where the conversion $1 \ {\rm cm}^2/{\rm g} =
1.8 \times 10^{-24} \ {\rm cm}^2/{\rm GeV}$ is used).  Although these
are just rough estimates, they show that the self-interactions of the
$\chi$ fermions are much smaller than the general range $0.1 \lsim
\sigma_{SI}/m_{DM} \lsim 1$ cm$^2$/g that is determined by fits to
cosmological observations (e.g., \cite{tulin_yu}) in the framework of
a self-interacting dark matter particle with mass $m_{DM}$.  Hence,
our dark matter particle is a WIMP with very small self-interactions.


\section{Conclusions}
\label{conclusion_section}

In this work we have constructed and studied a model for dark matter
with $n=2$ extra spatial dimensions in which Standard-Model 
fermions have localized wave functions. The underlying gauge group is
$G_{\rm SM} \otimes {\rm U}(1)_z$ and the dark matter particle is a
SM-singlet Dirac fermion, $\chi$, charged under the ${\rm U}(1)_z$
gauge symmetry.  The communication between the dark matter sector and
the SM fields arises mainly from the property that the SM fermions are
charged under the U(1)$_z$ gauge interaction.  This communication is
naturally weak, so $\chi$ is a weakly interacting massive particle,
which, furthermore, has negligibly small self-interactions.
We have focused on a subclass denoted $C_1$ in which a seesaw
mechanism is operative that naturally explains small masses for the
observed neutrinos. Within this subclass we have studied two versions
of the model, namely leptophilic and hadrophilic and have demonstrated
that each of these can account for cosmological observations on the
relic dark matter density. These fits allow a range of dark matter
masses, typically in the multi-TeV range.  Constraints from
direct-detection searches for dark matter can be relevant for the
hadrophilic version of the model, and we have determined the values of
the parameters that satisfy these constraints.  We also discuss
experimental tests of the model.

Separately from our application to dark matter, we have shown how the
model constitutes a counterexample to conventional wisdom in effective
field theory claiming that a Dirac fermion naturally has a mass lying
at the ultraviolet cutoff.  The reason that this conventional lore
does not apply here is due to the exponentially strong suppression of
the 4D Dirac fermion mass arising from the separate location of the
$\chi_L$ and $\chi_R$ wave function centers in the extra dimensions.
As a consequence of this suppression, for a moderate dimensionless distance
$\|\eta_{\chi_L}-\eta_{\chi_R}\|$ in the extra dimensions, the $\chi$ mass
parameter in the higher-dimensional space can be at the UV cutoff while
the physically observed $m_\chi$ is much smaller than this UV cutoff. 


\begin{acknowledgments}

 This research was supported in part by the U.S. National Science
 Foundation Grant NSF-PHY-1915093. We thank R. N. Mohapatra and
 S. Nussinov for valuable discussions on related work. 

\end{acknowledgments}


\bigskip
\bigskip

\begin{appendix}

\section{Neutral Vector Boson Masses and Mixing}
\label{vectorbosons}

In this appendix we briefly remark on neutral vector boson masses and
mixing in this model.  This will be treated in the context of the
low-energy 4D effective field theory. A covariant derivative acting on
the SM Higgs field $H$ is
\beq
D_\mu H = \Big ( {\mathbb I}_{2 \times 2}\partial_\mu - i g {\vec T} \cdot 
{\vec A}_\mu - \frac{g'Y_H}{2} B_\mu - \frac{g_z z_H}{2} C_\mu \Big )\, H \ ,
\label{DH}
\eeq
where ${\vec A}_\mu$, $B_\mu$, and $C_\mu$ are the gauge fields for the 
SU(2)$_L$, U(1)$_Y$, and U(1)$_z$ gauge interactions; 
$g$, $g'$, and $g_z$ are the respective gauge couplings, $Y_H=1$ is the 
weak hypercharge of
the SM Higgs $H$, and $z_H$ is the U(1)$_z$ charge of the $H$. A covariant
derivative acting on the additional Higgs field $\omega$ is 
\beq
D_\mu \omega = \Big ( \partial_\mu - i\frac{g_z}{2}C_\mu \Big )\, \omega
\ ,
\label{Domega}
\eeq
with the vacuum expectation values
\beq
\langle H \rangle_0 = \left( \begin{array}{c}
    0 \\
    \frac{v}{\sqrt{2}} \end{array} \right ) \ .
\label{hvev}
\eeq
and
\beq
\langle \omega \rangle_0 = \frac{v_\omega}{\sqrt{2}} \ , 
\label{omegavev}
\eeq
these yield the following mass-squared terms for the neutral gauge
fields in the 4D Lagrangian: 
\beqs
&& \frac{v^2}{8}(-gA^3_\mu + g'B_\mu + g_z z_H C_\mu)
             (-gA^{3 \mu} + g'B^\mu + g_z z_H C^\mu) \cr\cr
&& + \frac{g_z^2v_\omega^2}{8}C_\mu C^\mu \ .
\label{massquaredterms}
\eeqs
The resultant mixing has been analyzed in \cite{adh} and has the
property that in the limit $g_z v_\omega \gg z_H v$, the mixing
between $C_\mu$ and the SM gauge fields $A^3_\mu$ and $B_\mu$
vanishes.  Specifically, in this limit, the mixing angle vanishes like
$z_H (g^2+g'^2)^{1/2}v/(g_z v_\omega)$. Since we assume the above
inequality $g_z v_\omega \gg z_H v$, we neglect this mixing here.
(Recall that the SM factor $(g^2+g'^2)^{1/2} \simeq 0.74$ at a
reference scale $m_Z$, where $g$ and $g'$ are the running SU(2)$_L$
and U(1)$_Y$ gauge couplings; for a general discussion of effects
of heavy neutral vector bosons, see, e.g., \cite{pgl}.)
We also make use of the fact, as
discussed in \cite{adh,holdom}, that at a given scale one can perform a
rotation in the $(B_\mu,C_\mu)$ space to eliminate a kinetic mixing
term $\propto B_{\mu\nu}C^{\mu\nu}$ in the effective action,
so that it is only generated at loop level.

\end{appendix} 


\newpage



\begin{thebibliography}{99}
\bibitem{gennote}
Inserting the generational index $a$ we have, for example, 
$u_{1,R} \equiv u_R$, $u_{2,R} \equiv c_R$, $u_{3,R} \equiv t_R$;
$d_{1,R} \equiv d_R$, $d_{2,R} \equiv s_R$, $d_{3,R} \equiv b_R$;
$\ell_{1,R} \equiv e_R$, $\ell_{2,R} \equiv \mu_R$, $\ell_{3,R} \equiv \tau_R$,
etc.
\bibitem{dif}
Note that this model is quite different
from models in which only gravitons propagate in the extra dimensions
\cite{ged}, as is evident from the fact that the compactification scales
are quite different. For example, for $n=2$ and a quantum-gravity scale 
of 30 TeV, the compactification size in the models of Ref. \cite{ged} 
is $\sim 3 \times 10^{-4}$ cm, much larger than the scale
$L \simeq 2 \times 10^{-19}$ cm in our model.
\bibitem{steig2}
Parenthetically, it may be recalled that if the
dark matter were a self-conjugate particle, then the right-hand side of Eq.
(\ref{sigmav_needed}) would be reduced by half, to $2 \times 10^{-26}$ cm$^3$/s
\cite{steigman_dm}.)
\end{thebibliography}
\end{document}